\documentclass[a4paper,12pt]{article}
\usepackage{jheppub}
\usepackage{booktabs}
\usepackage{multirow} 
\newcommand{\be}{\begin{equation}}
	\newcommand{\ee}{\end{equation}}
\def\sfrac#1#2{{\textstyle{#1\over #2}}}
\newcommand{\bea}{\begin{eqnarray}}
	\newcommand{\eea}{\end{eqnarray}}
\newcommand{\nn}{\nonumber}

\newcommand{\HENs}{HE$\nu$s}
\newcommand{\DM}{{\scriptscriptstyle\rm DM}}

\title{High Energy Neutrinos, Gravitational Waves, and Dark Matter  from a Cosmological First Order Phase Transition}

\author[(1,3)]{James M. Cline,}
\emailAdd{jcline@physics.mcgill.ca}
\author[(2)]{Savas Stoica}
\emailAdd{savas.stoica@mail.mcgill.ca}
\author[(1)]{and Yong Xu}
\emailAdd{yong.xu6@mcgill.ca}
\affiliation[(1)]{McGill University Department of Physics \& Trottier Space Institute\\
	3600 Rue University, Montréal, QC, H3A 2T8, Canada}
\affiliation[(2)]{Oxford University Department of Physics, Clarendon Laboratory,
	Parks Road, Oxford, OX1 3PU}
\affiliation[(3)]{Niels Bohr International Academy,The Niels Bohr Institute,
	Blegdamsvej 17, DK-2100 Copenhagen Ø, Denmark}

\abstract{We investigate high energy neutrinos (\HENs) from the decay of superheavy dark matter (DM) produced during a cosmological first order phase transition. Adopting the filtered DM production mechanism, we revisit the relic abundance calculation by incorporating recent progress in determining the bubble wall velocity from hydrodynamics. We show that DM with mass $m_{\DM}\simeq 4.4\times10^{8}~\text{GeV}$ can reproduce the observed relic abundance and accommodate ultra high energy neutrino events such as KM3-230213A, while the same phase transition simultaneously generates a stochastic gravitational wave (GW) background. 
	Although the GW signal (which peaks at $10^{4}\,\text{Hz}$) is not yet observable in this example, 
	at lower DM masses $m_{\DM}\lesssim10^{6}~\text{GeV}$, the neutrino spectrum shifts into the energy range relevant for IceCube observations, while the associated GW signal moves to lower frequencies and can reach the projected sensitivities of Cosmic Explorer, BBO, and DECIGO. This gives a multimessenger connection between \HENs  ~ and GWs, with both signals connected to DM production in a  first order phase transition.}

\begin{document}
	
	\begin{flushright}
		\today
	\end{flushright}
	
	\maketitle

	\section{Introduction}
	
	First order phase transitions (FOPTs) in the early Universe provide a well defined setting in which physics beyond the Standard Model can leave observable cosmological signatures. During such transitions, bubble nucleation and expansion drive the plasma out of equilibrium and can generate a stochastic gravitational wave (GW) background. The resulting spectrum depends on quantities such as the released vacuum energy, the duration of the transition, and the bubble wall velocity, and therefore carries information about the underlying particle physics dynamics \cite{Weir:2017wfa,Athron:2023xlk}.
	
	High energy neutrino observations provide a complementary window on physics at energy scales difficult to access with terrestrial experiments. The IceCube experiment has observed astrophysical neutrinos extending to PeV energies \cite{IceCube:2013low}, including individual events such as IceCube-170922A \cite{IceCube:2018dnn}. More recently, KM3NeT reported KM3-230213A at substantially higher energy \cite{KM3NeT:2025npi}. These observations motivate the study of mechanisms that produce high energy neutrinos and, where possible, connect them to other probes of the particles' cosmological origin.
	
	Connections between GW signals and high energy neutrinos have been explored in light of interpretations of KM3-230213A. Examples include cosmic string scenarios, in which the GW background is generated by the string network while the neutrino flux arises from the decay of superheavy dark matter (DM) \cite{Khan:2025gxs,Barman:2025hoz}, and primordial black hole scenarios, where neutrinos are produced through heavy particle production or decay while GWs originate from the formation or subsequent evolution of the black holes \cite{Jiang:2025blz,Choi:2025hqt}. Related constructions involve domain wall dynamics as the source of GWs together with decaying DM as the origin of the neutrino signal \cite{Kohri:2025bsn}. See also Refs.~\cite{Murase:2025uwv,Borah:2025igh,Aloisio:2025nts} for interpretations of KM3-230213A based on decaying DM.
	
	In many of these scenarios, the source of GWs does not itself determine the abundance of the DM that produces the neutrinos. The connection between the two signals therefore depends on additional assumptions about DM production. A more direct link arises if the GW production event also determines the abundance of the DM that eventually decays into neutrinos.
	
	In this work, we investigate this possibility in the context of  cosmological first order phase transitions. We consider high energy neutrinos from the decay of superheavy DM, whose mass and relic abundance are generated during the transition. The same transition can also produce a stochastic GW background. The DM abundance, neutrino energies, and GW spectrum are thus connected through the properties of the phase transition, while the neutrino flux additionally depends on the DM lifetime and decay channels.
	
	We adopt the filtered DM production mechanism \cite{Baker:2019ndr,Chway:2019kft}, in which the DM particle is thermalized in the symmetric phase but acquires a mass that is larger than the temperature in the broken phase. Particles in the surrounding plasma can enter the expanding bubbles only if they carry sufficient energy to overcome the mass increase across the wall. The resulting transmission is suppressed, and the final relic abundance is determined by the fraction of the DM population that penetrates into the broken phase. This mechanism allows DM masses far above the conventional thermal scale while reproducing the observed relic abundance \cite{Baker:2019ndr,Chway:2019kft,Marfatia:2020bcs,Chao:2020adk,Jiang:2023nkj,Hosseini:2024ahe,Wada:2026iux}.
	
	The bubble wall velocity is an important quantity in this framework. It controls the flux of particles transmitted through the wall and hence enters directly in the determination of the DM abundance. At the same time, it affects the production and spectral shape of the GW background. Previous studies of filtered DM commonly treated the wall velocity as an adjustable parameter. Here we revisit the calculation by incorporating recent hydrodynamic results for the wall velocity \cite{Laurent:2022jrs,Ai:2023see,Ekstedt:2024fyq}, thus predicting the filtered abundance and GW spectrum within a coherent treatment of the phase transition dynamics.

	If the resulting heavy DM is metastable, its decay at late times can produce high energy neutrinos. We identify regions of parameter space in which the observed DM abundance is reproduced and the neutrino spectrum extends to energies relevant for KM3-230213A. This event provides an illustrative benchmark for our study. The corresponding phase transition produces a GW spectrum that typically peaks  around $ 10^{4}~\mathrm{Hz}$, with an amplitude below projected sensitivities. For lighter DM, the neutrino is within the energy range probed by IceCube, while the corresponding GW spectrum shifts toward lower frequencies and may become accessible to Cosmic Explorer, BBO, and DECIGO.
	
	The framework therefore relates three phenomena that are often considered separately: the abundance of superheavy DM, the high energy neutrino signal from its decay, and the GW background from the phase transition responsible for its production. The neutrino spectrum carries information about the DM mass and decay, while the GW spectrum probes the dynamics of the transition. Together, these observables offer a concrete multimessenger test of the phase transition origin of superheavy DM.  Moreover, the model we consider is directly related to mass generation of the light neutrinos by the seesaw mechanism.
	
	The remainder of the paper is organized as follows. In Sec.~\ref{sec:DM} we discuss DM production through the filtering mechanism. Sec.~\ref{sec:PT} describes the phase transition dynamics. In Sec.~\ref{sec:Neutrino} we compute the neutrino flux from DM decay. The GW spectra are shown in Sec.~\ref{sec:GW}. We summarize our conclusions in Sec.~\ref{sec:conclusion}.
	
	\section{Filtered Dark Matter Mechanism}\label{sec:DM}
	In our setup, the phase transition plays a crucial role in generating dark matter (DM) abundance. 
	In particular, we adopt the filter DM mechanism \cite{Baker:2019ndr, Chway:2019kft}. 
	We assume the DM candidate to be a right-handed neutrino (RHN) that couples to a scalar field $S$ undergoing a first-order phase transition. 
	The relevant interaction Lagrangian is
	\begin{align}\label{eq:yukawa_DM}
		\mathcal{L}_{\text{int.}} \supset \frac12
		y_s S \overline{N_i}^c N_i 
		+ y_{\nu,ji} \overline{L}_j \tilde H N_i 
		+ \text{h.c.}\,,
	\end{align}
	where $H$ is the Standard Model Higgs doublet, $L$ denotes the lepton doublet, and $N_i$ are several species of RHNs.  
	
	To generate the SM neutrino masses by the seesaw mechanism, at least two $N_i$ species are needed.  We assume there are three, with equal couplings to $S$.  Then without loss of generality, in the mass eigenbasis of the light neutrinos, there will be one linear combination $\tilde N$, which is a heavy mass eigenstate, coupling only to the lightest neutrino $\nu_1$ \cite{Cline:2020mdt}.  Denote its coupling to $\nu_1$ as $\tilde y_\nu$.  Since $m_{\nu_1}$ can be arbitrarily small, so can $\tilde y_\nu$, and this allows $\tilde N$ to be a long-lived DM candidate.  The decay rate of $\tilde N\to \nu_1 h$ can be written as
	\be
	\Gamma_{\tilde N} = {m_{\nu_1} M_{N}^2\over 16\pi\,v^2}
	\ee
	where $v=246$\,GeV is the Higgs VEV. For example, if the lifetime of $\tilde N$ is $10^{30}$\,s and $M_N\simeq 4.4\times 10^{8}~\text{GeV}$ (as suggested by the KM3NeT event), the corresponding light neutrino mass is $m_{\nu_1}\sim 10^{-56}$\,eV.  Chiral symmetry protects the small value of $\tilde y_\nu$ from quantum corrections.  The other two $N_i$ states decay quickly.

	Prior to the phase transition, the RHNs remain in thermal equilibrium with the plasma through interactions mediated by the scalar field $S$. 
	As the Universe cools, the scalar field undergoes a phase transition and develops a vacuum expectation value $v_s$, generating a large DM mass \begin{equation}
		m_{\DM}(T)=y_s\, v_s(T),
	\end{equation}
	where $v_s(T)$ is the temperature-dependent minimum of the potential.
	Inside bubbles of the true vacuum, where $S$ takes its nonzero value, the DM becomes heavy and effectively decouples from the thermal bath. 
	Outside the bubbles, in the false vacuum, the DM remains light and continues to annihilate efficiently. 
	As the bubbles expand and eventually percolate, only a small fraction of DM particles from the false vacuum can penetrate into the true vacuum because of the large mass gap across the bubble wall. 
	This selective transmission acts as a filter, resulting in a suppressed relic abundance determined by the bubble dynamics and phase-transition parameters \cite{Baker:2019ndr,Chway:2019kft}.
	
	The number density of DM transmitted into the broken phase is given by \cite{Chway:2019kft}
	\begin{align}\label{eq:nDM}
		n_{\DM,b} \simeq
		\frac{g_{\rm DM}\,T_n^{3}}{\gamma_w\,v_w}
		\left[
		\frac{\gamma_w(1-v_w)\,m_{\DM}/T_n+1}
		{4\pi^{2}\gamma_w^{3}(1-v_w)^{2}}
		\right]
		\exp \left[-\frac{\gamma_w(1-v_w)\,m_{\DM}}{T_n}\right]\,,
	\end{align}
	where $g_{\rm DM}$ is the number of internal degrees of freedom of DM, $T_n$ corresponds to the nucleation temperature,   $v_w$ is the bubble wall velocity, and $\gamma_w = (1-v_w^2)^{-1/2}$ is the associated Lorentz factor. Throughout this work, we take $g_{\rm DM}=2$, appropriate for a Majorana neutrino. The corresponding DM yield is
	\begin{align}\label{eq:YDM}
		Y_{\DM} = \frac{ n_{\DM,b}}{s}\,, \quad
		s = \frac{2\pi^2}{45}\, g_{\star s}\, T^3\,,
	\end{align}
	where  $s$
	is the entropy density, with $g_{\star s}$ denoting the effective number of entropy degrees of freedom. After the phase transition, and in the absence of further number-changing processes or entropy injection, $Y_{\DM}$ approaches a constant and can be identified with the late-time comoving DM abundance.
	
	The observed DM density is given by
	\cite{Planck:2018vyg,ParticleDataGroup:2024cfk}
	\begin{align}\label{eq:mY}
		m_{\DM} \,Y_{\DM} \simeq 4.3 \times 10^{-10} ~\text{GeV}\,,
	\end{align}
	where $m_{\DM} = y_s v_{s,{T=0}} \equiv  y_s v_0$ denotes the zero-temperature DM mass. In our analysis, Eq.~\eqref{eq:YDM} is used to determine the relic abundance, while the phase-transition parameters, including $T_n$, are obtained in the next section from the phase-transition dynamics. It is important to notice that the DM abundance depends sensitively on the bubble wall velocity, $v_w$, as shown in Eq.~\eqref{eq:nDM}. In most previous studies, $v_w$ was assumed to be a free parameter. In this work, we adopt the recent hydrodynamic treatment of Ref.~\cite{Ai:2023see}, which provides a significant improvement over the approximations used in Refs.~\cite{Baker:2019ndr,Chway:2019kft}.

	\section{Phase Transition Dynamics}
	\label{sec:PT}
	
	We consider a generic scalar potential at finite temperature that leads to a first-order phase transition \cite{Dine:1992wr,Quiros:1999jp,Kehayias:2009tn, Wang:2020jrd},
	\begin{align}
		V(S,T)
		=
		\frac{\mu^2+DT^2}{2}S^2
		-\xi T S^3
		+\frac{\lambda_s}{4}S^4 .
		\label{eq:potential}
	\end{align}
	It corresponds to the leading terms in the high-temperature expansion, including the cubic term that would be induced by integrating out bosonic 
	particles that get mass only from the VEV $\langle S\rangle$: $\xi = \sum_b g_b(m_b/S)^3/(12\,\pi)$.  The thermal mass term $DT^2$ on the other hand receives contributions from both scalars and fermions that get their masses from $\langle S\rangle$.

	In the context of filtered DM,  $D$ cannot be made arbitrarily small, because it gets contributions both from $S$ itself and from the unspecified light degrees of freedom that contribute to the cubic term.\footnote{In principle, there are also contributions from $N_i$, but these are highly suppressed inside the bubbles because of $M_N\gg T$.}  From the high-$T$ expansion
	of the one-loop finite-temperature potential \cite{Quiros:1999jp,Laine:2016hma}, the quadratic contributions are  
	\bea
	\Delta V_T
	&=&
	\sum_bg_b {m_b^2(S)T^2\over 24}\nn\\
	&=&
	\left( g_\xi^{1/3}{(12\pi\xi)^{2/3}\over 24} + {\lambda_s \over 8}\right) T^2 S^2,
	\eea
	where $g_{b}$ counts bosonic internal degrees of freedom, and $g_\xi$ is specific to the massless boson (in the symmetric phase) that contributes to $\xi$.  Comparing with $V(S,T)\supset \frac12DT^2S^2$, 
	and assuming that $g_\xi=1$, one has 
	\begin{equation}
		D\geq D_{\min}
		=
		\left(\pi^2\xi^2\over 12\right)^{1/3} + {\lambda_s \over 4}\,.
		\label{eq:thermal_mass_cut}
	\end{equation}
	This condition prevents taking the limit in which the thermal restoring force is made
	arbitrarily small, and therefore removes one part of parameter space that could otherwise
	strengthen the transition.  For the parameters of interest that we will find, the second term in
	Eq.\ (\ref{eq:thermal_mass_cut}) is always negligible.
	
	A second limitation on the smallness of $D$ is that there must be a phase transition.  Algebraically this corresponds to the requirement that the critical temperature, at which the symmetric and broken minima are degenerate, is real-valued.  From
	Eq.~\eqref{eq:potential}, it follows that
	\begin{equation}
		T_c=
		\frac{\lambda_s v_0}{\sqrt{\lambda_sD-2\xi^2}},
	\end{equation} 
	so that $\lambda_sD>2\xi^2$. It is therefore useful to define the  parameter
	\begin{equation}
		\epsilon
		\equiv
		\frac{2\xi^2}{\lambda_sD},
		\qquad
		0<\epsilon<1\, ,
		\label{eq:epsilon_def}
	\end{equation}
	which measures how close the potential is to the  boundary of real-valued $T_c$'s; for $\epsilon>1$, the symmetric phase is always the true vacuum.
	Small $\epsilon$ corresponds to a shallow tunneling barrier, while $\epsilon$ close to unity
	corresponds to a large barrier near the edge of the allowed region.
	
	The quantity controlling the overall scale  of $v_s(T)$ in units of temperature
	is $\xi/\lambda_s$.  Defining
	\begin{equation}
		\delta(T)
		\equiv
		\frac{\lambda_s(\mu^2+DT^2)}{\xi^2T^2},
	\end{equation}
	the VEV can be written as
	\begin{equation}
		\frac{v_s(T)}{T}
		=
		\frac{3\xi}{2\lambda_s}
		\left[
		1+\sqrt{1-\frac{4\delta(T)}{9}}
		\right].
		\label{eq:vs_over_T}
	\end{equation}
	In the nucleation region of interest, the bracket is of order unity.  Thus at the nucleation temperature (defined below), 
	$v_n/T_n$ is mainly determined by $\xi/\lambda_s$.
	At fixed $D$ and $\epsilon$, the cubic coefficient is
	$\xi=(\epsilon\lambda_sD/2)^{1/2}$, and therefore
	\begin{equation}
		\frac{\xi}{\lambda_s}
		=
		\sqrt{\frac{\epsilon D}{2\lambda_s}} \,.
		\label{eq:q_from_epsilon}
	\end{equation}
	Decreasing $\lambda_s$ increases $\xi/\lambda_s$ and hence increases the tunneling
	barrier.  This is the expected direction in parameter space for decreasing the
	transmitted yield  and increasing $m_{\DM}$.  However, at the same time,   it increases the 
	strength of
	the phase transition, leading to higher wall velocities, possibly running away to $v_w=1$,
	which will invalidate the filtering mechanism.
	
	The phase transition occurs when the rate of bubble nucleation per Hubble volume equals the Hubble rate. 
	The phase transition proceeds through thermal tunneling from the false vacuum to
	the true vacuum.  The nucleation temperature $T_n$ is the temperature at which
	bubbles of the broken phase form efficiently.  We determine it by the condition
	that approximately one bubble is nucleated per Hubble volume \cite{Quiros:1999jp} 
	\begin{align}\label{eq:S3T_number}
		\frac{S_3(T_n)}{T_n}\simeq 137 + \ln \left(\frac{10^2 \xi^2}{\lambda D/2}\right) + 4 \ln \left(\frac{100~\text{GeV}}{T_n}\right),
	\end{align}
	where $S_3(T)$ is the three dimensional Euclidean action of the critical bubble. The right hand side of Eq.~\eqref{eq:S3T_number} reduces to the familiar number $\sim 140$ for a phase transition at electroweak scale.
	We evaluate $S_3(T)/T$ using the analytic approximation of
	Ref.~\cite{Adams:1993zs},
	\begin{align}
		\frac{S_3}{T}
		=
		\frac{64\sqrt{2}\pi}{81}
		\frac{\xi}{\lambda_s^{3/2}}
		(2-\delta)^{-2}
		\left(\beta_1\delta+\beta_2\delta^2+\beta_3\delta^3\right)
		\sqrt{\frac{\delta}{2}},
		\label{eq:S3T}
	\end{align}
	where $\delta$ is defined above and
	\begin{align}
		\beta_1=8.2983,
		\qquad
		\beta_2=-5.5330,
		\qquad
		\beta_3=0.8180 .
	\end{align}
	The approximation Eq.~\eqref{eq:S3T} provides an accurate description of the tunneling action in
	the parameter range relevant for our numerical scan, as we verified using the numerical package CosmoTransitions \cite{Wainwright:2011kj}.
	
	Apart from $T_n$, the transition is characterized by the thermodynamic
	quantities entering the solution for the bubble wall properties.  The strength of the phase transition is
	quantified by~\cite{Hindmarsh:2020hop,Athron:2023xlk}
	\begin{align}\label{eq:alpha_n}
		\alpha_n
		=
		\left.
		\frac{\left(1-\frac{1}{4}T\frac{\partial}{\partial T}\right)\Delta V}
		{\rho_R(T)}
		\right|_{T=T_n},
	\end{align}
	where
	\begin{equation}
		\Delta V\equiv V(0,T)-V(v_s,T) > 0
	\end{equation}
	is the vacuum energy difference between the symmetric and broken phases, and
	$\rho_R(T)=\frac{\pi^2}{30}g_\star(T) T^4
	$ is the radiation energy density.  Here $g_\star (T)$ denotes the total number of
	relativistic degrees of freedom contributing to the radiation energy density in the broken phase; we take account of its $T$-dependence in our scans.
	$\alpha_n$ represents the released vacuum energy relative to the
	background radiation energy density.
	The inverse duration of the transition is denoted by
	\begin{align}
		\frac{\beta}{H_n}
		=
		\left.
		\frac{d(S_3/T)}{d\ln T}
		\right|_{T=T_n},
	\end{align}
	where $H_n$ is the Hubble rate at $T_n$; it governs how 
	the nucleation rate depends on temperature.  Smaller values of $\beta/H_n$
	correspond to a longer-lasting transition.
	
	To compute the wall velocity using the method of Ref.~\cite{Ai:2023see}, we
	need the ratio of enthalpy densities\footnote{See Appendix \ref{app:thermo_wall} for  details.} across the bubble wall at $T_n$,
	\begin{align}
		\Psi_n \equiv \frac{w_b(T_n)}{w_s(T_n)}\, ,
	\end{align}
	where $w\equiv\rho+p$ is the enthalpy density, with $\rho$ and $p$ the energy
	density and pressure, respectively.  The subscripts $b$ and $s$ denote the
	broken and symmetric phases.  The quantities $\alpha_n$, $\beta/H_n$, and
	$\Psi_n$ are the inputs that connect the scalar potential to the hydrodynamic solution for the bubble wall.  In the numerical scan described below, they determine whether a
	given phase transition has a subluminal wall velocity, and its value if so.  One can show that  nucleation can only take place if  $   \alpha_n > \sfrac13(1-\Psi_n)
	\label{alphacond}$
	since otherwise $\Delta V<0$ and there is no tunneling to the broken phase. Throughout the analysis we assume the sound speed $c_s^2=1/3$ on both sides of the wall.
	
	\subsection{Scale Invariance of Bounce Action and Parameter Scan Strategy}
	\label{massfixing}
	In this work, we wish to explain the high energy  neutrino event where  neutrinos arise from the decay of DM.  The DM mass $m_{\DM}$ is then fixed to be the energy scale of the neutrino event. On the other hand, $m_{\DM} Y_{\DM}$ is also fixed to reproduce the observed DM relic abundance as shown in Eq.~\eqref{eq:mY}. Thereafter, the task reduces to search for the parameter space for a fixed $Y_{\DM}$ for a given  $m_{\DM}$.

	Since the DM yield $Y_{\DM}$ in Eq.~\eqref{eq:YDM} is dimensionless, it is useful to express the action in terms of a dimensionless field $s$, temperature $\tau$ and coordinate $\hat x$ by  introducing 
	\begin{equation}
		s\equiv \frac{S}{v_0},
		\qquad
		\tau\equiv \frac{T}{v_0}\,\qquad \hat x \equiv v_0\, x\,.
	\end{equation}
	Eq.~\eqref{eq:potential} becomes
	\begin{equation}
		V(S,T)
		=
		v_0^4
		\left[
		\frac12(-\lambda_s+D\tau^2)s^2
		-\xi\tau s^3
		+\frac{\lambda_s}{4}s^4
		\right] \equiv v_0^4 \hat V(s,\tau)\,,
		\label{eq:scaled_potential}
	\end{equation}
	and the 3D high-temperature action is
	\be
	S_3 = v_0\int d^{\,3}\hat x\left(\sfrac12s'^2 + \hat V\right) \equiv v_0\,\hat S_3
	\ee
	where $s' = ds/d\hat r$, the derivative with 
	respect to the dimensionless radial coordinate.
	Hence $S_3/T = \hat S_3/\tau$, and all reference to the overall scale $v_0$ has disappeared.
	
	It follows that $T_c/v_0$, $T_n/v_0$, $v_n/v_0$, and hence $v_n/T_n$ are fixed
	by the dimensionless  potential $\hat V$.  The same is true for the
	dimensionless thermodynamic quantities entering the hydrodynamic wall solution,
	including $\alpha_n$, $\beta/H_n$, $\Psi_n$ and $v_w$.  Therefore the
	transmitted yield is a function of the dimensionless parameters only,
	\begin{equation}
		Y_{\DM}=Y_{\DM}(\theta),
		\qquad
		\theta=(\lambda_s,D,\epsilon,y_s).
	\end{equation}
	
	In the following, we will scan over the parameter space to search for $Y_{\DM}$, and for a given DM mass:
	\begin{equation}
		m_{\DM}
		=
		\frac{4.3\times10^{-10}\ {\rm GeV}}
		{Y_{\DM}({\theta\in{\cal V}})} ,
		\label{eq:mDM_res}
	\end{equation}
	where ${\cal V}$ is the set of parameters satisfying thermal consistency,
	nucleation, and the subluminal wall velocity condition specified in previous section.  For a Majorana fermion coupled as
	$\mathcal{L}\supset -y_s S\bar{N}^cN/2$, perturbative unitarity implies
	$y_s\leq\sqrt{4\pi}$ \cite{Allwicher:2021rtd}, which leads to the lowest transition scale (typically most promising for observations) at 
	a fixed DM mass.
	Lowering $y_s$ increases the symmetry-breaking scale $v_0$
	required to reproduce the same fermion mass.  It does not change
	the dimensionless analysis of the filtering mechanism or the qualitative behavior discussed
	below, although quantities such as the nucleation temperature and
	the resulting GW frequencies scale with $1/y_s$. {In our numerical analysis, we examine several values of
		$y_s$ to assess this scale dependence and identify the parameter choice
		that offers the most optimistic prospects for GW detection while satisfying
		all relevant constraints.}

	\section{High Energy Neutrino Flux}\label{sec:Neutrino}
	
	\begin{table*}[t]
		\centering
		\resizebox{\textwidth}{!}{%
			\begin{tabular}{ccccccccc}
				\toprule
				$D/D_{\min}$
				& Point
				& $\lambda_s$
				& $\epsilon$
				& $T_n~[\mathrm{GeV}]$
				& $\alpha$
				& $\beta/H_n$
				& $v_w$
				& $Y_{\DM}/Y_{\DM}^{\mathrm{target}}$
				\\
				\midrule
				
				\multirow{3}{*}{$1$}
				& 1 & $3.352\times10^{-5}$ & $0.9990$
				& $1.195\times10^{7}$ & $1.198\times10^{-2}$
				& $9.635\times10^{3}$ & $0.59352$ & $1.040962$
				\\
				& 2 & $3.765\times10^{-5}$ & $0.9824$
				& $1.236\times10^{7}$ & $1.157\times10^{-2}$
				& $9.166\times10^{3}$ & $0.56572$ & $1.058917$
				\\
				& 3 & $4.229\times10^{-5}$ & $0.9713$
				& $1.277\times10^{7}$ & $1.128\times10^{-2}$
				& $8.696\times10^{3}$ & $0.53367$ & $0.787981$
				\\
				\midrule
				
				\multirow{3}{*}{$5$}
				& 1 & $2.420\times10^{-4}$ & $0.4555$
				& $7.101\times10^{6}$ & $3.117\times10^{-1}$
				& $1.227\times10^{4}$ & $0.66522$ & $0.974827$
				\\
				& 2 & $2.420\times10^{-4}$ & $0.4611$
				& $7.080\times10^{6}$ & $3.193\times10^{-1}$
				& $1.219\times10^{4}$ & $0.67141$ & $1.053892$
				\\
				& 3 & $2.154\times10^{-4}$ & $0.4999$
				& $6.740\times10^{6}$ & $3.758\times10^{-1}$
				& $1.237\times10^{4}$ & $0.71839$ & $1.062519$
				\\
				\midrule
				
				\multirow{3}{*}{$10$}
				& 1 & $4.329\times10^{-4}$ & $0.6330$
				& $4.507\times10^{6}$ & $4.808$
				& $1.095\times10^{4}$ & $0.89108$ & $1.023080$
				\\
				& 2 & $1.520\times10^{-4}$ & $0.6441$
				& $3.439\times10^{6}$ & $5.050$
				& $1.827\times10^{4}$ & $0.93577$ & $0.950943$
				\\
				& 3 & $2.719\times10^{-4}$ & $0.6385$
				& $3.995\times10^{6}$ & $4.928$
				& $1.373\times10^{4}$ & $0.91443$ & $1.124800$
				\\
				\midrule
				
				\multirow{3}{*}{$15$}
				& 1 & $5.337\times10^{-5}$ & $0.6607$
				& $1.934\times10^{6}$ & $18.14$
				& $3.741\times10^{4}$ & $0.97989$ & $1.146034$
				\\
				& 2 & $4.751\times10^{-5}$ & $0.6607$
				& $1.879\times10^{6}$ & $18.15$
				& $3.966\times10^{4}$ & $0.98024$ & $0.518302$
				\\
				& 3 & $5.995\times10^{-5}$ & $0.6607$
				& $1.992\times10^{6}$ & $18.12$
				& $3.529\times10^{4}$ & $0.97952$ & $2.453462$
				\\
				\bottomrule
			\end{tabular}%
		}
		\caption{
			Benchmark parameter points with DM yields closest to the target value
			$Y_{\DM}^{\text{target}}\simeq 9.8\times10^{-19}$, for different
			values of $D/D_{\min}$. The DM mass is fixed at
			$m_{\DM}\simeq 4.4\times10^{8}~\mathrm{GeV}$, motivated by the
			energy of KM3-230213A.
		}
		\label{tab:benchmarks_PT}
	\end{table*}
	
	Our dark matter candidate is a heavy sterile neutrino $ N$ with mass $m_{\DM}$ and lifetime $\tau_{\DM}$, which decays producing Standard Model (SM) neutrinos, both directly and through the decay of the Higgs boson in $N\to h\nu$. The differential neutrino flux received at Earth has two contributions: $(i)$ from the Milky Way (MW) and $(ii)$ from the extragalactic (EG) cosmological DM.
	In the two-body decay $N\to \nu h$,  the energy spectrum of neutrino from the primary decay is  monochromatic with energy $E \simeq {m_{\DM}}/{2}$. To account for the secondary neutrino flux from the intermediate Higgs decay, we have utilized the  package \texttt{HDMSpectra} \cite{Bauer:2020jay}.

	Integrating along a line of sight (l.o.s.) at an angle $\theta$ relative to the Galactic center, the flux per unit solid angle is
	\begin{equation}\label{eq:MWflux}
		\frac{d^2\Phi^{\text{dec.}}_{\rm MW}}{dE\,d\Omega}(\theta)
		= \frac{1}{4\pi} \int_{\rm l.o.s.} \frac{\rho_{\rm DM}(r(\ell,\theta))}{m_{\DM} \tau_{\DM}}\, \frac{dN_\nu}{dE}\, d\ell
		= \frac{1}{4\pi}\, \frac{1}{m_{\DM} \tau_{\DM}}\, \frac{dN_\nu}{dE} \, D(\theta),
	\end{equation}
	where $dN_\nu/dE$ is the number of neutrinos produced per $N$ decay per unit energy in the source rest frame, and
	\begin{equation}
		D(\theta) \equiv \int_{\rm l.o.s.} \rho_{\rm DM}(r(\ell,\theta))\, d\ell
	\end{equation}
	is the line-of-sight integral (D-factor), with units $\mathrm{GeV\,cm^{-2}}$. For an NFW halo normalized to
	$\rho_\odot=0.4~{\rm GeV\,cm^{-3}}$, the line of sight toward KM3-230213A
	gives $D\simeq 10^{22}~{\rm GeV\,cm^{-2}}$.  For the isotropic extragalactic component, the observed differential flux per unit solid angle is
	\begin{equation}\label{eq:EGflux}
		\frac{d^2\Phi^{\text{dec.}}_{\rm EG}}{dE\,d\Omega} = \frac{c}{4\pi} \frac{\Omega_{\rm DM} \rho_c}{m_{\DM} \tau_{\DM}} \int_0^\infty \frac{dz}{(1+z) H_0 \sqrt{\Omega_m (1+z)^3 + \Omega_\Lambda}} \frac{dN_\nu}{dE'}\,,
	\end{equation}
	where $E'=E/(1+z)$ is the neutrino energy at redshift $z$.  This describes the broadening of the intrinsic line spectrum by the expansion of the Universe. The total differential flux from the galaxy and from the Universe is \begin{align}
		\frac{d^2\Phi^{\text{dec.}}}{dE\,d\Omega} \equiv \frac{d^2\Phi^{\text{dec.}}_{\rm MW}}{dE\,d\Omega} + \frac{d^2\Phi^{\text{dec.}}_{\rm EG}}{dE\,d\Omega}\,.
	\end{align} 
	
	\begin{figure}[t!]
		\def\sepf{0.42}
		\centering
		\includegraphics[scale=\sepf]{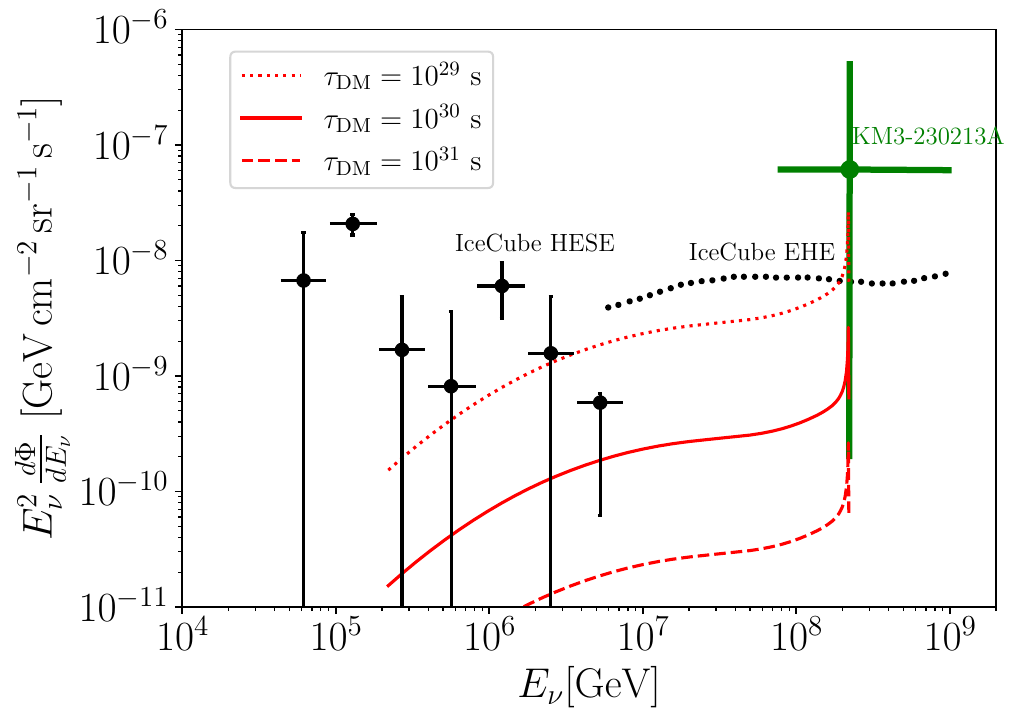}
		\includegraphics[scale=\sepf]{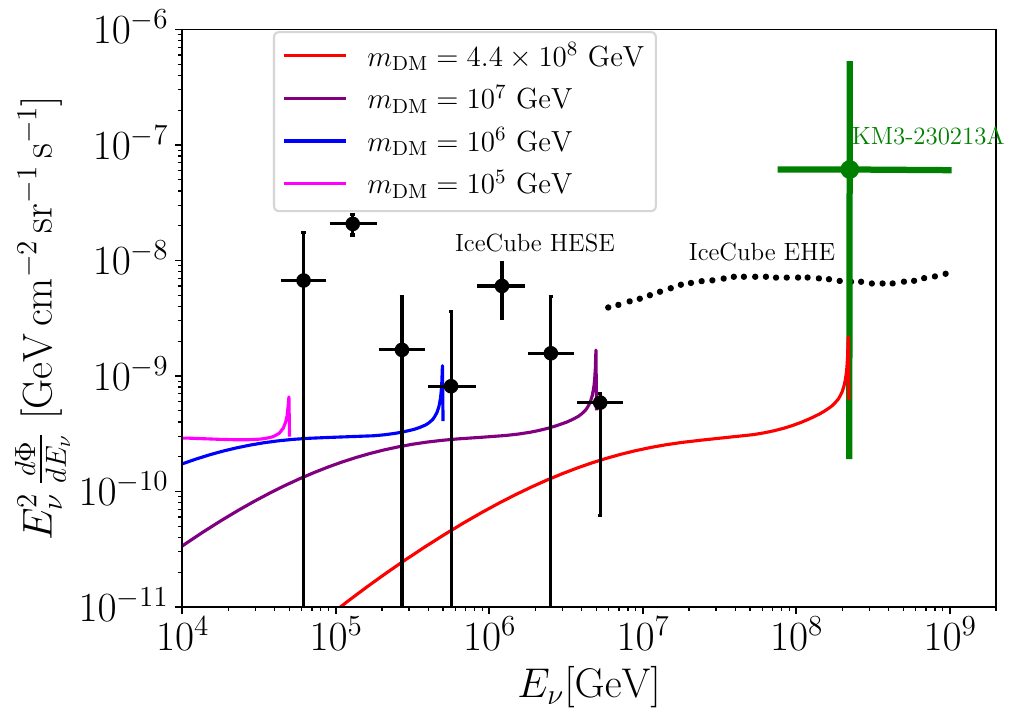}
		\caption{SM neutrino flux from DM decays for different DM lifetimes $\tau_{\DM}$ at $m_{\DM} \simeq 4.4\times 10^{8}~\text{GeV}$ (left), and for different DM masses with $\tau_{\DM} =10^{30} ~\text{s}$ (right), using the benchmark parameters listed in Table~\ref{tab:benchmarks_PT}. 
			The black data points correspond to IceCube EHE~\cite{IceCube:2018fhm} and IceCube HESE~\cite{IceCube:2020wum}, while the green data point corresponds to KM3-230213A~\cite{KM3NeT:2025npi}.}
		\label{fig:flux}
	\end{figure} 

	In the left frame of Fig.~\ref{fig:flux}, we show the neutrino flux from DM decay for  $m_{\DM} \simeq 4.4\times 10^{8}\,\text{GeV}$ with the corresponding model parameters given in Table~\ref{tab:benchmarks_PT} {and assuming $y_s = \sqrt{4\pi}$. The results presented in Fig.~\ref{fig:flux} are independent of this choice.} The red dotted, solid, and dashed curves correspond to DM lifetimes of $\tau_{\DM}=10^{29}\,\text{s}$, $10^{30}\,\text{s}$, and $10^{31}\,\text{s}$, respectively. A shorter lifetime implies a larger decay rate and hence a larger neutrino flux. 
	We find that for DM lifetimes $\tau_{\rm DM}\gtrsim 5 \times  10^{29}\,\mathrm{s}$, the predicted neutrino flux can account for KM3-230213A~\cite{KM3NeT:2025npi} while remaining consistent with the data from IceCube EHE in 2018~\cite{IceCube:2018fhm}. In the right panel, the red, purple, blue, and magenta solid curves correspond to $m_{\DM} \simeq 4.4\times 10^{8}\,\text{GeV}$, $10^{7}\,\text{GeV}$, $10^{6}\,\text{GeV}$, and $10^{5}\,\text{GeV}$, respectively. As the DM mass decreases, the neutrino spectrum shifts toward lower energies, making it potentially relevant for the IceCube HESE data~\cite{IceCube:2020wum}.
	
	Here  we mainly focus on the continuous neutrino spectrum generated by Higgs boson decay and the subsequent electroweak showering\footnote{Electroweak showering also produces an accompanying gamma-ray component at high dark matter masses, which can provide complementary constraints on decaying dark matter~\cite{Arguelles:2022nbl}. Since the lifetimes considered here are consistent with these bounds, we focus on the neutrino and gravitational wave signals.}, which we compute using \texttt{HDMSpectra} \cite{Bauer:2020jay}. In the two body decay channel $N\to h\nu$, the primary neutrino additionally gives a monochromatic contribution at the kinematic endpoint, $E\simeq m_{\rm DM}/2$. Since this contribution is concentrated within a very narrow energy interval (broadened only by the velocity distribution of the DM), its significance is reduced by the experimental energy resolution. 
	We therefore focus here on the continuous component, which provides the more extended spectral contribution relevant for comparison with the presently available high energy neutrino data. Given the sizable experimental uncertainties, including the localized endpoint contribution is not expected to qualitatively alter the comparison shown in Fig.~\ref{fig:flux}.

	\section{Gravitational Wave Spectrum}\label{sec:GW}
	
	A first order phase transition can generate a stochastic gravitational wave
	(GW) background through the dynamics of the scalar field and the surrounding
	thermal plasma. The main contributions arise from acoustic motion of the
	plasma, magnetohydrodynamic turbulence and, when a nonnegligible fraction of
	the released vacuum energy remains stored in the scalar field configuration,
	collisions of the expanding bubbles
	\cite{Hindmarsh:2015qta,Caprini:2015zlo,Caprini:2019egz,Caprini:2024hue}.
	These contribute to the GW spectrum as
	\begin{equation}
		\Omega_{\rm GW} h^2
		=
		\Omega_{\rm sw} h^2
		+
		\Omega_{\rm turb} h^2
		+
		\Omega_{\rm coll} h^2 ,
		\label{eq:GW_total}
	\end{equation}
	where $\Omega_{\rm sw}$, $\Omega_{\rm turb}$, and $\Omega_{\rm coll}$
	denote the contributions from sound waves, turbulence, and scalar field
	bubble collisions, respectively. 
	
	\paragraph{Sound waves.}
	
	Once the expanding bubbles transfer a significant fraction of the released
	vacuum energy to the surrounding plasma, the resulting bulk fluid motion
	continues after the bubbles collide and merge. The corresponding acoustic
	waves provide an efficient source of gravitational radiation and can give
	the dominant contribution for thermal phase transitions in which the bubble
	walls do not run away
	\cite{Hindmarsh:2015qta,Caprini:2019egz}.
	
	We parametrize this contribution
	using the  simulation-based spectral fit of Ref.~\cite{Caprini:2015zlo}
	for the sound-wave spectrum, supplemented by the finite-lifetime
	suppression factor derived in Ref.~\cite{Guo:2020grp}.
	The resulting empirical fit is \cite{Caprini:2015zlo,Guo:2020grp,Luo:2025ewp}
	\begin{equation}
		\Omega_{\rm sw} h^2
		\simeq
		2.65\times10^{-6}\,
		\left(\frac{H_n}{\beta}\right)
		\left(
		\frac{\kappa_{\rm sw}\alpha}{1+\alpha}
		\right)^2
		\left(\frac{100}{g_*}\right)^{1/3}
		\Upsilon_{\rm sw}\,v_w\,S_{\rm sw}(f).
		\label{eq:Omega_sw}
	\end{equation}
	The efficiency factor $\kappa_{\rm sw}$ is the fraction
	of the released vacuum energy converted into bulk kinetic energy of the
	plasma. For strong phase transition with large $\alpha$ and sizable wall velocity, $\kappa_{\rm sw} \to 1$ \cite{Caprini:2015zlo}.  The factor $\Upsilon_{\rm sw}$ accounts for the finite duration
	of the acoustic source and will be defined below. The frequency dependence of the acoustic contribution is given by
	the fit \cite{Caprini:2015zlo,Hindmarsh:2017gnf}
	\begin{equation}
		S_{\rm sw}(f)
		=
		\left(\frac{f}{f_{\rm sw}}\right)^3
		\left[
		\frac{7}
		{4+3\left(f/f_{\rm sw}\right)^2}
		\right]^{7/2},
		\label{eq:S_sw}
	\end{equation}
	where $f$ is the GW frequency observed today and $f_{\rm sw}$ is the
	peak frequency. For a transition occurring during radiation domination,
	$ f_{\rm sw}$ is approximately \cite{Caprini:2015zlo}
	\begin{equation}
		f_{\rm sw}
		\simeq
		1.9\times10^{-5}\,{\rm Hz}\,
		\frac{1}{v_w}
		\left(\frac{\beta}{H_n}\right)
		\left(\frac{T_n}{100~{\rm GeV}}\right)
		\left(\frac{g_*}{100}\right)^{1/6}.
		\label{eq:f_sw}
	\end{equation}
	
	The acoustic source does not  remain active indefinitely.
	As the fluid perturbations evolve, nonlinear effects lead to shock
	formation and eventually destroy the coherent acoustic motion. The
	corresponding sound wave lifetime $\tau_{\rm sw}$ can be estimated from the time required
	for fluid perturbations to propagate over the characteristic bubble
	separation \cite{Ellis:2020awk,Caprini:2019egz},
	\begin{equation}
		\tau_{\rm sw}
		\simeq
		\frac{R_*}{\overline U_f},
		\qquad
		R_*
		\simeq
		(8\pi)^{1/3}\frac{v_w}{\beta},
		\qquad
		\overline U_f^{\,2}
		\simeq
		\frac{3}{4}
		\frac{\kappa_{\rm sw}\alpha}{1+\alpha},
		\label{eq:tau_sw}
	\end{equation}
	where $R_*$ is the mean bubble separation and $\overline U_f$ is the root-mean-square fluid velocity. The last relation follows from the kinetic
	energy carried by a relativistic plasma
	\cite{Hindmarsh:2017gnf,Caprini:2019egz}.
	
	For a radiation-dominated Universe, the finite lifetime of the acoustic
	source modifies the GW amplitude through the suppression factor derived
	in Ref.~\cite{Guo:2020grp},
	\begin{equation}
		\Upsilon_{\rm sw}
		=
		1-
		\frac{1}
		{\sqrt{1+2H_s\tau_{\rm sw}}}\,,
		\label{eq:Upsilon_sw}
	\end{equation}
	where $H_s$ is the Hubble parameter at the onset of the acoustic
	source. For a sufficiently fast phase transition characterized by
	$\beta/H_n \gg 1$, the Hubble rate changes only negligibly over the
	transition timescale, so that $H_s \simeq H_n$ is a good
	approximation.
	Physically, $\Upsilon_{\rm sw}$ in Eq.~\eqref{eq:Upsilon_sw} measures how much of the GW signal has
	time to build up before the coherent acoustic motion is disrupted. If the
	source survives for only a small fraction of a Hubble time,
	$H_n\tau_{\rm sw}\ll1$, the GW production is correspondingly reduced,
	with $\Upsilon_{\rm sw}\simeq H_n\tau_{\rm sw}$. In the opposite limit,
	$H_n\tau_{\rm sw}\gg1$, the acoustic source persists sufficiently long
	for the GW signal to approach the long-lived limit, and
	$\Upsilon_{\rm sw}\rightarrow1$.

	\paragraph{Hydrodynamic turbulence.}
	
	The nonlinear evolution of the plasma can transfer part of its kinetic
	energy into vortical and magnetohydrodynamic turbulence. Although the
	normalization and detailed spectral shape of this contribution remain more
	uncertain than those of the acoustic component, it provides an additional
	source of GWs and is conventionally included in estimates of the total
	signal \cite{Caprini:2009yp,Caprini:2015zlo,Caprini:2019egz}. We use the
	standard empirical form \cite{Caprini:2015zlo,Caprini:2019egz} 
	\begin{equation}
		\Omega_{\rm turb} h^2
		\simeq
		3.35\times10^{-4}
		\left(\frac{H_n}{\beta}\right)
		\left(
		\frac{\kappa_{\rm turb}\alpha}{1+\alpha}
		\right)^{3/2}
		\left(\frac{100}{g_*}\right)^{1/3}
		v_w\,
		S_{\rm turb}(f),
		\label{eq:Omega_turb}
	\end{equation}
	where $\kappa_{\rm turb}$ is the fraction of the released vacuum energy that
	eventually sources turbulent plasma motion. It is conventionally related to the
	acoustic efficiency factor through $\kappa_{\rm turb}
	=
	\epsilon_{\rm turb}\,\kappa_{\rm sw},$
	where $\epsilon_{\rm turb}$ parametrizes the fraction of the bulk fluid
	energy transferred into turbulence. Following the commonly adopted  choice of
	Ref.~\cite{Caprini:2015zlo}, we take
	$\epsilon_{\rm turb}=0.05$, although its precise value is uncertain
	\cite{Caprini:2019egz}.
	
	The spectral shape of the turbulent contribution is parametrized as \cite{Caprini:2009yp,Caprini:2015zlo,Caprini:2019egz}
	\begin{equation}
		S_{\rm turb}(f)
		=
		\frac{
			\left(f/f_{\rm turb}\right)^3
		}{
			\left(1+f/f_{\rm turb}\right)^{11/3}
			\left(1+8\pi f/h_*\right)
		},
		\label{eq:S_turb}
	\end{equation}
	where the characteristic frequency redshifted to today is
	\begin{equation}
		f_{\rm turb}
		\simeq
		2.7\times10^{-5}\,{\rm Hz}\,
		\frac{1}{v_w}
		\left(\frac{\beta}{H_n}\right)
		\left(\frac{T_n}{100~{\rm GeV}}\right)
		\left(\frac{g_*}{100}\right)^{1/6},
		\label{eq:f_turb}
	\end{equation}
	and
	\begin{equation}
		h_*
		\simeq
		1.65\times10^{-5}\,{\rm Hz}\,
		\left(\frac{T_n}{100~{\rm GeV}}\right)
		\left(\frac{g_*}{100}\right)^{1/6}
		\label{eq:hstar}
	\end{equation}
	is the Hubble frequency at the transition redshifted to the present epoch. The factor involving
	$h_*$ in Eq.~\eqref{eq:S_turb} accounts for the finite decorrelation time of
	the turbulent source.

	\paragraph{Bubble collisions.}
	
	A further contribution can arise directly from the scalar field gradients
	associated with the expanding bubble walls. It is most
	important when an appreciable fraction of the released vacuum energy remains
	stored in the scalar field configuration up to the time of bubble collisions.
	The associated efficiency factor is denoted by $\kappa_\phi$, the fraction of the released energy retained
	in the scalar field. For walls
	that rapidly reach a terminal velocity because of their interaction with the
	thermal plasma, $\kappa_\phi$ is generally suppressed, whereas it can become
	important for sufficiently strong transitions with highly relativistic or
	runaway walls
	\cite{Huber:2008hg,Caprini:2015zlo,Caprini:2019egz}.
	
	Estimating the scalar field contribution using the
	standard envelope approximation gives \cite{Huber:2008hg,Caprini:2015zlo}
	\begin{equation}
		\Omega_{\rm coll} h^2
		\simeq
		1.67\times10^{-5}
		\left(\frac{H_n}{\beta}\right)^2
		\left(
		\frac{\kappa_\phi\alpha}{1+\alpha}
		\right)^2
		\left(\frac{100}{g_*}\right)^{1/3}
		\frac{0.11\,v_w^3}{0.42+v_w^2}
		S_{\rm coll}(f)\,.
		\label{eq:Omega_coll}
	\end{equation}
	Within this approximation, the
	spectral shape can be fitted by \cite{Huber:2008hg,Caprini:2015zlo}
	\begin{equation}
		S_{\rm coll}(f)
		=
		\frac{
			3.8\left(f/f_{\rm coll}\right)^{2.8}
		}{
			1+2.8\left(f/f_{\rm coll}\right)^{3.8}
		}\,,
		\label{eq:S_coll}
	\end{equation}
	with the corresponding present day peak frequency
	\begin{equation}
		f_{\rm coll}
		\simeq
		1.65\times10^{-5}\,{\rm Hz}\,\left(
		\frac{0.62}
		{1.8-0.1v_w+v_w^2}\right)
		\left(\frac{\beta}{H_n}\right)
		\left(\frac{T_n}{100~{\rm GeV}}\right)
		\left(\frac{g_*}{100}\right)^{1/6}.
		\label{eq:f_coll}
	\end{equation}
	Both expressions are obtained from numerical simulations of bubble
	collisions within the envelope approximation.

	For the benchmark points considered in this work, the finite bubble wall velocities imply substantial energy transfer to the plasma, so that the sound wave contribution provides the dominant component over most of the relevant parameter space. The turbulent contribution is generally subleading but represents an additional plasma source and is therefore retained. By contrast, the bubble collision contribution becomes important primarily when a sizable fraction of the released vacuum energy remains stored in the scalar field configuration of the expanding walls, as can occur for sufficiently strong transitions with walls approaching the relativistic limit. In our numerical analysis, the wall velocities remain appreciably below the speed of light, indicating that most of the released energy is transferred to the plasma rather than retained in the scalar field sector. We therefore neglect the collision contribution and include the sound wave and turbulence terms in Eq.~\eqref{eq:GW_total}. This treatment is consistent with the hydrodynamic regime realized by our benchmark points and captures the dominant GW sources throughout the parameter space studied here.

	\begin{figure}[t]
		\def\sepf{0.7}
		\centering
		\includegraphics[scale=\sepf]{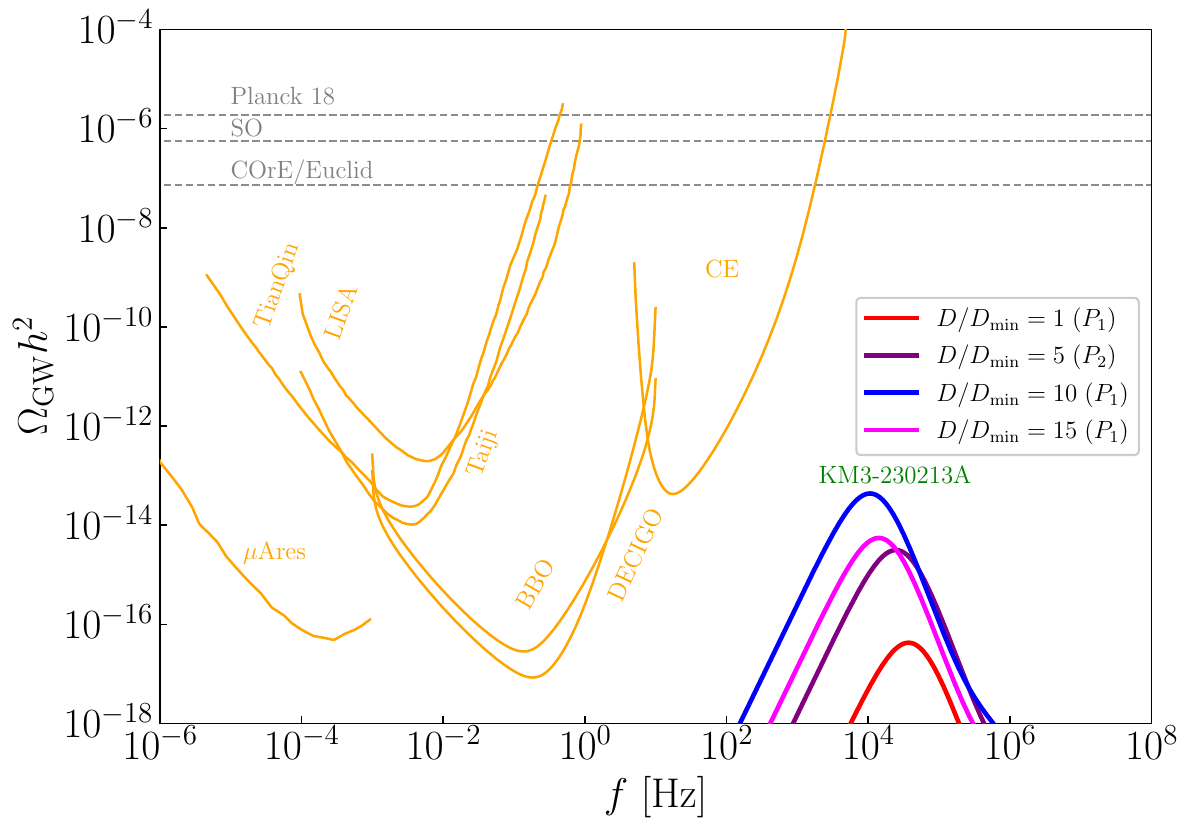}
		\caption{
			GW spectra for benchmark points motivated by the energy of KM3-230213A, with $m_{\DM}\simeq 4.4\times10^{8}~\mathrm{GeV}$, {and the thermal potential parameter $D$ varying around its optimum range for producing GWs}. 
			Dashed lines are current (Planck 2018) and projected constraints on $N_{\rm eff}$.
			The target yield corresponding to the observed DM relic abundance is given in Eq.\ (\ref{eq:mY}). 
			The model parameters are listed in Table~\ref{tab:benchmarks_PT}.
		}
		\label{fig:GW}
	\end{figure} 
	
	Using the phase transition parameters obtained in the previous section, we compute the GW spectra for representative benchmark points.  The resulting spectra are shown in Fig.~\ref{fig:GW}. Here, we also overlay projected sensitivity curves (shown in yellow) for the Big Bang Observer (BBO)~\cite{Crowder:2005nr, Corbin:2005ny}, ultimate DECIGO~\cite{Seto:2001qf, Kudoh:2005as}, LISA~\cite{LISA:2017pwj}, TianQin~\cite{TianQin:2015yph}, Taiji~\cite{Ruan:2018tsw}, $\mu$Ares~\cite{Sesana:2019vho}, Cosmic Explorer (CE)~\cite{Reitze:2019iox}, and the Einstein Telescope (ET)~\cite{Hild:2010id, Punturo:2010zz, Sathyaprakash:2012jk, ET:2019dnz}. Moreover, the energy density stored in gravitational waves redshifts as radiation and therefore contributes to the effective number of relativistic degrees of freedom, $N_{\text{eff}}$. 
	Constraint on $N_{\text{eff}}$ are shown as horizontal dashed red lines in Fig.~\ref{fig:GW}. 
	The Planck 2018 results give $N_{\text{eff}} = 2.99 \pm 0.34$ at $95\%$ confidence level~\cite{Planck:2018vyg}. 
	Future surveys such as COrE~\cite{COrE:2011bfs} and Euclid~\cite{EUCLID:2011zbd} are expected to reach sensitivities of $\Delta N_{\text{eff}} \lesssim 0.013$ at the $2\sigma$ level, while the next-generation CMB experiment CMB-S4 is projected to achieve $\Delta N_{\text{eff}} \lesssim 0.06$~\cite{Abazajian:2019eic}.
	
	In Fig.~\ref{fig:GW}  each curve corresponds to one of the parameter choices listed in Table~\ref{tab:benchmarks_PT}, corresponding to $m_{\DM} \simeq 4.4\times 10^{8}\,\text{GeV}$.
	The bubble wall velocity is determined using the hydrodynamic method of Ref.~\cite{Ai:2023see}, ensuring consistency between the thermodynamic parameters and the dynamical evolution of the transition.
	As shown in Fig.~\ref{fig:GW}, increasing $D$ from its minimum value initially enhances the overall
	GW amplitude and then suppresses it, with the largest signal produced
	around $D\simeq 10D_{\min}$ with $\Omega_{\text{GW}} h^2 \simeq 2 \times 10^{-14}$. The turnover of the peak amplitude can be
	understood from the approximate scaling $\Omega_{\rm GW}^{\rm peak}h^2
	\propto
	\frac{v_w}{(\beta/H_n)^2}
	\left(\frac{\alpha}{1+\alpha}\right)^2$.
	At smaller $D/D_{\min}$, the growth of $\alpha$ enhances the GW signal.
	Once the strength factor $\alpha/(1+\alpha)$ begins to saturate, however,
	the increase of $\beta/H_n$ at larger $D/D_{\min}$ dominates and suppresses
	the peak amplitude.
	As shown in Table~\ref{tab:benchmarks_PT}, the nucleation temperatures
	required for the stronger transitions with larger $\alpha$ are typically
	of order $T_n\sim10^{6}~{\rm GeV}$. Together with the corresponding large
	values of $\beta/H_n$, this places the GW peak at frequencies of order
	$10^{4}~{\rm Hz}$, consistent with Eq.~\eqref{eq:f_sw}. {These numerical results were obtained with $y_s = \sqrt{4\pi}$. Keeping the DM mass fixed, decreasing $y_s$
		raises the characteristic scale of the phase transition, shifting the GW
		peak toward higher frequencies and reducing its amplitude. The GW spectra presented therefore represent an optimistic limiting case,
		illustrating the optimal observational reach of the framework while
		satisfying all relevant constraints.}
	
	\begin{figure}[t]
		\def\sepf{0.7}
		\centering
		\includegraphics[scale=\sepf]{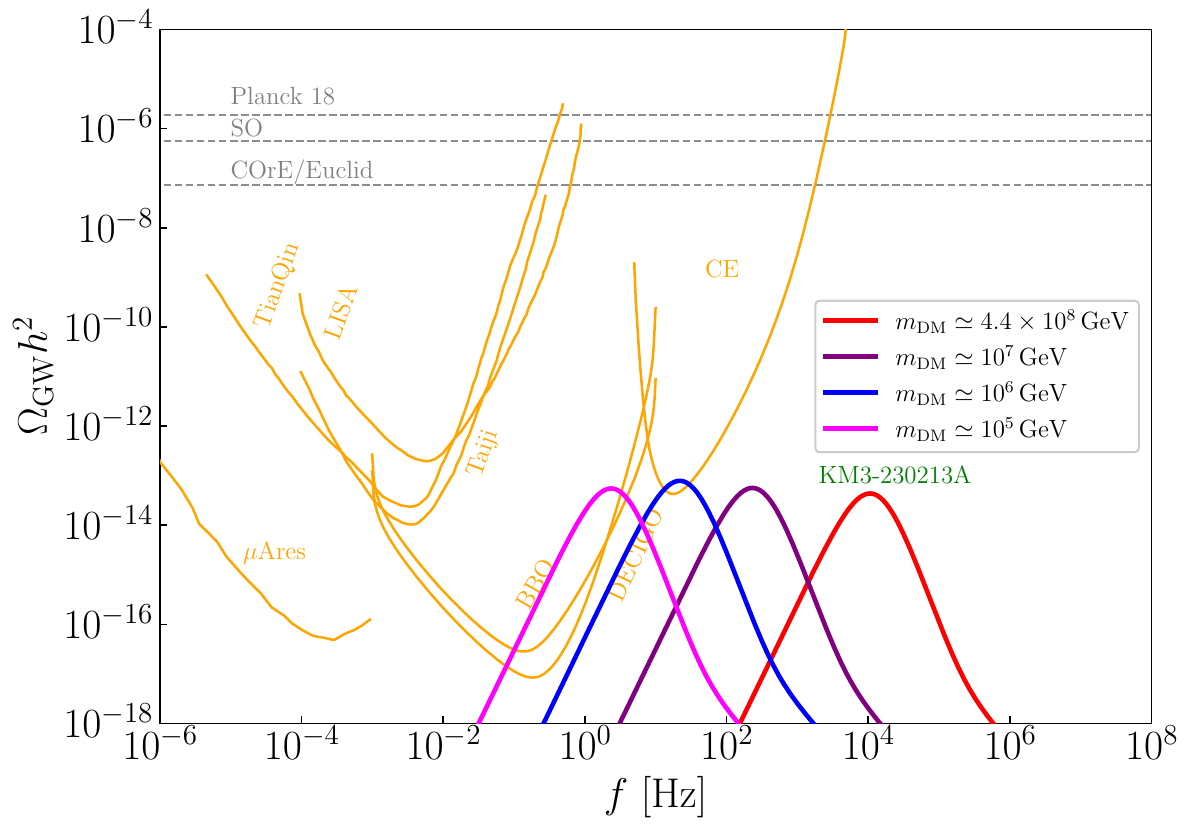}
		\caption{The corresponding GW spectrum for four benchmark DM masses, the corresponding model parameters are listed in
			Table~\ref{tab:benchmarks_PT2}.}
		\label{fig:GW2}
	\end{figure} 
	
	\begin{table*}[t]
		\centering
		\resizebox{\textwidth}{!}{%
			\begin{tabular}{cccccccc}
				\toprule
				$\lambda_s$
				& $\epsilon$
				& $T_n~[\mathrm{GeV}]$
				& $\alpha$
				& $\beta/H_n$
				& $v_w$
				& $Y_{\DM}/Y_{\DM}^{\mathrm{target}}$
				\\
				\midrule
				$4.329\times10^{-4}$ & $0.6330$
				& $4.507\times10^{6}$ & $4.808$
				& $1.095\times10^{4}$ & $0.89108$ & $1.023080$ \\
				$6.136\times10^{-4}$ & $0.6358$
				& $1.121\times10^{5}$ & $4.791$
				& $9.653\times10^{3}$ & $0.89364$ & $0.8875315$ \\
				$9.326\times10^{-4}$ & $0.6358$
				& $1.250\times10^{4}$ & $4.720$
				& $8.075\times10^{3}$ & $0.88440$ & $0.7632674$ \\
				$6.136\times10^{-4}$ & $0.6448$
				& $1.119\times10^{3}$ & $4.897$
				& $1.009\times10^{4}$ & $0.92071$ & $0.9759281$ \\
				\bottomrule
			\end{tabular}%
		}
		\caption{Benchmark points for four representative DM masses:
			$m_{\DM}\simeq 4.4\times10^{8}\,\mathrm{GeV}$,
			$10^{7}\,\mathrm{GeV}$, $10^{6}\,\mathrm{GeV}$, and
			$10^{5}\,\mathrm{GeV}$. All points have $D/D_{\min}=10$ and $y_s=\sqrt{4\pi}$.}
		\label{tab:benchmarks_PT2}
	\end{table*}

	In Fig.~\ref{fig:GW2}, we show the predicted GW spectra for $m_{\DM} \simeq 4.4\times 10^{8}\,\text{GeV}$, $10^{7}\,\text{GeV}$, $10^{6}\,\text{GeV}$, and $10^{5}\,\text{GeV}$, with the corresponding model parameters listed in Table~\ref{tab:benchmarks_PT2}. A clear correlation emerges between the DM mass and the characteristic GW frequency: as $m_{\DM}$ decreases, the scale of the phase transition is lowered, shifting the GW peak toward lower frequencies. Notably, for $m_{\DM}\lesssim 10^{6}\,\text{GeV}$, the predicted signals enter the projected sensitivity ranges of CE, BBO, and DECIGO. This provides a direct GW counterpart to the neutrino signal in Fig.~\ref{fig:flux}.
	
	\section{Conclusions}\label{sec:conclusion}
	
	In this work, we have studied a multimessenger realization of filtered dark matter produced during a cosmological first order phase transition.  The dark matter is assumed to get most of its mass from condensation of a scalar field in the phase transition, allowing its mass to exceed  the unitarity bound for thermal relic production. The mass gain of the DM particles in the true vacuum Boltzmann-suppresses their density, allowing them to have the observed relic abundance, and the phase transition generates a stochastic gravitational wave (GW) background, while the subsequent decay of dark matter can produce high energy neutrinos.
	
	A crucial aspect of our predictions is accurately estimating the bubble wall velocity $v_w$, which controls the filtering efficiency and thereby the final dark matter abundance. In previous studies, $v_w$ was treated as an adjustable parameter, or estimated crudely. Here we determine it from the hydrodynamics of the phase transition using recent methodological advances \cite{Laurent:2022jrs,Ai:2023see,Ekstedt:2024fyq}. This results in a constrained connection between the predicted abundance and GW signal.
	
	For $m_{\DM}\simeq 4.4\times10^{8}~\mathrm{GeV}$, we find parameter regions that reproduce the observed relic abundance and yield a neutrino flux at energies relevant for the high-energy KM3NeT event KM3-230213A, as shown in Fig.~\ref{fig:flux}. For DM with correspondingly large masses, {the GW spectrum  can peak}  around $10^{4}~\mathrm{Hz}$, with amplitudes reaching $\mathcal{O}(10^{-14})$, as shown in Fig.~\ref{fig:GW}.  These are below the sensitivity of currently planned experiments.  
	
	For producing observable GWs, we 
	considered lower dark matter masses, for which the neutrino spectrum is in the range of energies probed by IceCube. The corresponding phase transition occurs at at lower temperatures, shifting the GW spectrum toward lower frequencies. For $m_{\DM}\lesssim10^{6}~\mathrm{GeV}$, the predicted GW signal can reach the projected sensitivities of Cosmic Explorer, BBO, and DECIGO, as shown in Fig.~\ref{fig:GW2}.
	
	The connection between the possible neutrino and GW signals arises from the 
	dependence of the relic density on properties of the phase transition.  The observation of correlated signals would be suggestive of the framework we have explored, but probably not conclusive.  There is a further possible test through searches for the scale of absolute neutrino masses: our DM candidate is only long-lived if the lightest neutrino is essentially massless.  The combination of cosmology, oscillation experiments and searches for neutrinoless double beta decay may eventually give a lower bound on $m_{\nu_1}$ that could falsify the model.

	\begin{acknowledgments} 
		We thank Benoit Laurent for helpful correspondence  regarding  the wall velocity, and Andrew Long for discussions about the filtering mechanism.  JC thanks the Niels Bohr International Academy for hospitality during the completion of this work.  We  are supported by the Natural Sciences and Engineering Research Council (NSERC) of Canada.
	\end{acknowledgments}
	
	\appendix

	\section{Thermodynamic Quantities Entering the Finite Wall Solution}
	\label{app:thermo_wall}
	
	The wall velocity used in the main text is computed with the local thermal
	equilibrium framework of Ref.~\cite{Ai:2023see}.  In this framework, the
	existence and velocity of a steady finite wall solution are determined by
	thermodynamic quantities evaluated in the two phases of the transition.  The
	relevant inputs are the phase transition strength, the enthalpy ratio, and the sound
	speeds.  In this appendix we specify the enthalpy ratio used in the numerical
	scan.
	
	We denote the symmetric phase by the subscript $s$ and the broken phase by the
	subscript $b$.  The nucleation temperature is denoted by $T_n$.  The pressure,
	energy density, and enthalpy density of a phase are denoted by $p$, $\rho$, and
	$w$, respectively, with $w=\rho+p$.  The enthalpy ratio entering the finite wall
	solution is
	\begin{equation}
		\Psi_n
		\equiv
		\frac{w_b(T_n)}{w_s(T_n)} .
		\label{eq:Psi_def_app}
	\end{equation}
	Both $w_b$ and $w_s$ are evaluated at the same temperature $T_n$, as required by
	the thermodynamic matching used in Ref.~\cite{Ai:2023see}.
	
	In the symmetric phase, we approximate the thermodynamics by a relativistic
	radiation bath.  The quantity $g_s$ denotes the effective number of relativistic
	degrees of freedom in the symmetric phase.  With
	$\rho_s=\pi^2 g_sT^4/30$ and $p_s=\rho_s/3$, the symmetric phase enthalpy is
	\begin{equation}
		w_s(T)
		=
		\frac{4\pi^2}{90}g_sT^4 .
		\label{eq:ws_app}
	\end{equation}
	
	In the broken phase, the free energy contains a scalar vacuum contribution and a
	radiation contribution.  We denote by $g_b$ the effective number of
	relativistic degrees of freedom in the broken phase.  The broken phase field
	value at temperature $T$ is denoted by $v_s(T)$, and the scalar potential
	evaluated at this minimum is denoted by $V_b(T)\equiv V(v_s(T),T)$.  The broken
	phase free energy is therefore $F_b(T)=V_b(T)-\pi^2g_bT^4/90$.  The pressure is
	$p_b=-F_b$, and the energy density is $\rho_b=F_b-T\,dF_b/dT$.  Combining these
	relations gives the broken phase enthalpy
	\begin{equation}
		w_b(T)
		=
		- T\frac{dV_b(T)}{dT}
		+
		\frac{4\pi^2}{90}g_bT^4 .
		\label{eq:wb_general_app}
	\end{equation}
	Thus only the temperature dependence of the vacuum free energy contributes to
	the vacuum part of the enthalpy; a temperature independent vacuum energy
	contributes to $\rho$ and $p$ separately, but cancels in $w=\rho+p$.

	Using $V_b(T)=V(v_s(T),T)$, one has $dV_b/dT=DTv_s^2(T)-\xi v_s^3(T)$.  Equation~\eqref{eq:wb_general_app}
	then becomes
	\begin{equation}
		w_b(T)
		=
		\frac{4\pi^2}{90}g_bT^4
		-
		T\left[
		D\,T\,v_s^2(T)
		-
		\xi\,v_s^3(T)
		\right].
		\label{eq:wb_toy_app}
	\end{equation}
	
	Substituting Eqs.~\eqref{eq:ws_app} and \eqref{eq:wb_toy_app} into
	Eq.~\eqref{eq:Psi_def_app}, the enthalpy ratio used in the scan is
	\begin{equation}
		\Psi_n
		=
		\frac{
			\displaystyle
			\frac{4\pi^2}{90}g_bT_n^4
			-
			T_n\left[
			D\,T_n\,v_s^2(T_n)
			-
			\xi\,v_s^3(T_n)
			\right]
		}{
			\displaystyle
			\frac{4\pi^2}{90}g_sT_n^4
		}\,.
		\label{eq:Psi_final_app}
	\end{equation}
	A small value of $\Psi_n$ means that the enthalpy carried by the broken phase is
	much smaller than that of the symmetric phase.  In the finite wall solutions of
	Ref.~\cite{Ai:2023see}, this occurs as the transition becomes increasingly
	vacuum dominated and approaches the relativistic finite wall boundary.
	In the numerical analysis we take 
	\begin{equation}
		g_s = g_{\star} (T_n)+ 1 + 3\times 2\times\frac78,
		\qquad
		g_b = g_{\star} (T_n)+ 1.
		\label{eq:gs_gb}
	\end{equation}
	The number $g_{\star} (T_n)$ denotes the Standard Model relativistic degrees of freedom at $T_n$,
	while the extra $+1$ counts the real scalar $S$.  
	The term
	$3\times 2\times7/8$ is present only in the symmetric phase: there all the three right handed neutrinos are  light
	thermal Majorana fermion with $g=2$, and fermionic relativistic degrees of
	freedom carry the usual $7/8$ weight. In the broken phase right handed neutrino is heavy and
	its filtered abundance is Boltzmann suppressed, so it is not counted as part of
	the relativistic plasma.

	\bibliographystyle{JHEP} 
	\bibliography{biblio}
\end{document}